\documentclass[aps,pre,floatfix]{revtex4}
\usepackage{epsf}
\usepackage{subfigure}
\usepackage{amsmath}
\usepackage{amssymb}
\usepackage{graphicx}
\usepackage{epstopdf}
\begin{document}

\newcommand{\be}{\begin{equation}}
\newcommand{\ee}{\end{equation}}
\newcommand{\bea}{\begin{eqnarray}}
\newcommand{\eea}{\end{eqnarray}}
\newcommand{\mean}[1]{\left \langle #1 \right \rangle}

\title{\bf Broken ${\mathbb Z}_2$ symmetries and fluctuations in statistical mechanics}

\author{Pierre Gaspard}
\affiliation{Center for Nonlinear Phenomena and Complex Systems \& Department of Physics,\\
Universit\'e Libre de Bruxelles, Code Postal 231, Campus Plaine, 
B-1050 Brussels, Belgium}

\begin{abstract}
An analogy is developed between the breaking of space-inversion symmetry in equilibrium statistical mechanics and the breaking of time-reversal symmetry in nonequilibrium statistical mechanics.  In this way, similar relationships characterizing fluctuations are obtained in both contexts.
\end{abstract}

\noindent {}

\vskip 0.5 cm

\maketitle

\section{Introduction}
\label{Intro}

Symmetry breaking is a major phenomenon in natural systems. {\it A priori}, the equations describing a system may be symmetric under a group of continuous or discrete transformations.  For some reasons, the system is often found in some state that is a solution of the describing equations without satisfying all the symmetries of these equations.  In such cases, the symmetry is broken by the particular state of the system. Symmetry breaking phenomena may have different origins such as external forces or constraints  imposed to the bulk or the boundaries of the system.  In other cases, the symmetry may be broken by the initial conditions that are selected.  In spontaneous symmetry breaking, the symmetric state is unstable so that the system has to evolve towards some stable but asymmetric state, which is selected by slightly biased external perturbations or initial conditions.

These different phenomena can be studied for the symmetries that are broken.  They can be continuous leading to the emergence of Goldstone-Nambu massless modes (e.g., the transverse acoustic modes in solids) or Anderson-Brout-Englert-Higgs massive modes if a local gauge symmetry is broken in the presence of the corresponding long-range interaction.

Here, we focus on discrete ${\mathbb Z}_2$ symmetries such that the symmetry group is composed of the identity $1$ and an involution $X$, ${\mathbb Z}_2=\{1,X\}$.  Our purpose is to compare the breaking of time-reversal symmetry by thermodynamic forces in nonequilibrium steady states to the breaking of space inversion by external fields in equilibrium states.  In the later case, we aim at obtaining fluctuation relations similar to those known for nonequilibrium steady states, using the analogy between them through the phenomenon of symmetry breaking.

The paper is organized as follows.  In Section \ref{Hamilton}, the analogy between broken space-inversion and time-reversal symmetries is illustrated in simple Hamiltonian systems with one degree of freedom or infinitely many noninteracting particles.  In Section \ref{Space}, fluctuation relations are deduced for equilibrium spin systems in an external magnetic field breaking a space-inversion symmetry.  In Section \ref{Time}, nonequilibrium fluctuation relations are presented in analogy with the previous ones, emphasizing the role of the time-reversal symmetry breaking by the invariant probability measure describing the nonequilibrium steady state.  Conclusions and perspectives are drawn in Section \ref{Conclusions}.

\section{Symmetry breaking by selection of initial conditions}
\label{Hamilton}

In this section, we start by considering two Hamiltonian systems with one degree of freedom illustrating ${\mathbb Z}_2$ symmetry breaking.  The first is the Hamiltonian flow in a double-well potential and the second is the simple pendulum.  Both Hamiltonian systems are symmetric under space inversion $x\to - x$ (i.e., parity) and time reversal $t\to - t$.  However, there exist orbits breaking the symmetry under parity in the first system and the symmetry under time reversal in the second system.  As a consequence, the invariant probability measures supported by these orbits also break the corresponding symmetry.  In these one-dimensional systems, the orbits are periodic except the separatrices formed by the homoclinic orbits connected to the unstable stationary points.  For the periodic orbits, an invariant measure can be defined by temporal averaging over the period $T$ to get the invariant probability density
\be
\rho_ C(\Gamma) = \frac{1}{T} \int_0^T \delta(\Gamma-\Phi^t\Gamma_0) \, dt
\label{rho_time}
\ee
where $\Gamma=(x,p)$ denotes a point in phase space, $\Gamma_0$ is the initial condition of the orbit, and $\Phi^t$ is the Hamiltonian flow.  The support of the so-defined probability measure is the curve depicting the periodic orbit in the phase space:
\be
 C = \{ \Gamma = \Phi^t\Gamma_0 : \; t\in [0,T[ \} \; .
\ee

We may wonder if the invariant density (\ref{rho_time}) coincides with the microcanonical invariant density defined by
\be
\rho_E(\Gamma) = \frac{\delta\left[ E- H(\Gamma)\right]}{\int\delta\left[ E- H(\Gamma)\right]\, d\Gamma}
\label{rho_micro}
\ee
where $H(\Gamma)=H(x,p)$ is the Hamiltonian function and $E=H(\Gamma_0)$ is the energy fixed by the initial condition $\Gamma_0$.

Let us look at this question in the two aforementioned examples.

\vskip 0.3 cm

\noindent{\bf 1. Double-well potential.} 
For this system, the Hamiltonian function is given by
\be
H=\frac{p^2}{2m} + U(x) \qquad \mbox{where} \qquad U(x) = -\frac{a}{2} \, x^2 + \frac{b}{4} \, x^4
\label{dble-well}
\ee
with $a>0$ and $b>0$.  The phase space is the plane ${\mathcal M}=\{\Gamma=(x,p)\in{\mathbb R}^2\}$. The Hamiltonian function is invariant under the parity transformation:
\be
\Pi \, (x,p)=(-x,-p)
\ee
which is an involution since $\Pi^2=1$, generating the group ${\mathbb Z}_2=\{1, \Pi\}$ of symmetries.  Consequently, Hamilton's equations of motion stay invariant under parity.

\begin{figure}[h]
\begin{center}
\includegraphics[scale=0.35]{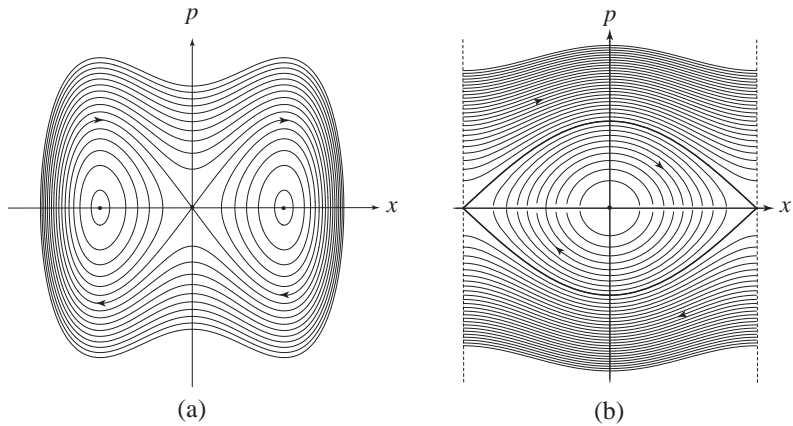}
\caption{Phase portraits of the Hamiltonian flows for (a) the double-well potential (\ref{dble-well}) and (b) the simple pendulum (\ref{pendulum}).}
\label{fig1}
\end{center}
\end{figure}

The phase portrait of this system is plotted in Fig.~\ref{fig1}a.  The system has three stationary points: the origin $x=p=0$ which is symmetric but unstable and the two minima of the potential at $x=\pm\sqrt{a/b}$ and $p=0$, which are stable at the minimum energy $U_{\rm min}=-a^2/(4b)$.  At zero energy, there are two homoclinic loops connected to the unstable point and they play the role of separatrices between the periodic orbits at negative and positive energies. 

With every positive value of the energy $E>0$, there corresponds one and only one periodic orbit $ C_E$ that oscillates between both wells.  In contrast, two periodic orbits $ C_E^{(\pm)}$ correspond with every negative value of the energy $U_{\rm min}<E<0$ and they oscillate in {\it either} the well on the right-hand side {\it or} the one on the left-hand side.  If the periodic orbits at positive energies are symmetric under parity, this is no longer the case for those at negative energies:
\bea
&&\Pi \,  C_E =  C_E \qquad\qquad\qquad\ \; \mbox{for}\qquad E>0 \label{C_E}\; ,\\
&&\Pi \,  C_E^{(+)} =  C_E^{(-)}\neq  C_E^{(+)} \qquad \mbox{for}\qquad E<0 \label{C_E+-}\; .
\eea
Therefore, every periodic orbit at negative energy breaks the symmetry under space inversion of Hamilton's equations.

\vskip 0.3 cm

\noindent{\bf 2. Simple pendulum.}
The Hamiltonian function is here given by
\be
H=\frac{p^2}{2m l^2 } -mgl  \cos x
\label{pendulum}
\ee
where $m$ is the mass of the body attached to a massless rod of length $l$ in the gravitational acceleration $g$ oriented towards $x=0$.  The angle varies in the interval $-\pi < x \leq +\pi$ so that the phase space is the cylinder ${\mathcal M}=\{\Gamma=(x,p): \; x\in]-\pi,+\pi], p\in{\mathbb R}\}$.
The Hamiltonian function and Hamilton's equations are symmetric under the time-reversal transformation:
\be
\Theta\, (x,p) = (x,-p)
\ee
which is also an involution since $\Theta^2=1$, generating the group ${\mathbb Z}_2=\{1, \Theta\}$ of symmetries.  

The phase portrait of the pendulum is plotted in Fig.~\ref{fig1}b.  The system has two stationary configurations: the stable one $x=p=0$ at the minimum energy $U_{\rm min}=-mgl$ and the unstable one $(x=\pm\pi,p=0)$ at the energy $U_{\rm max}=+mgl$.  On the cylinder, the two unstable points $(x=\pm\pi,p=0)$ are equivalent so that the two curves connecting them form two homoclinic orbits.  They constitute the separatrices between the oscillating periodic orbits at lower energies $U_{\rm min}<E<U_{\rm max}$ and the rotating periodic orbits at higher energies $E>U_{\rm max}$.

With every value of the energy $U_{\rm min}<E<U_{\rm max}$, there corresponds one and only one oscillating periodic orbit $ C_E$ which is self-reversed because the time-reversal transformation maps it onto itself.  In contrast, there exist two periodic orbits $ C_E^{(\pm)}$ rotating either anticlockwise or clockwise corresponding with every energy value $E>U_{\rm max}$.  None of them is self-reversed so that each one breaks the time-reversal symmetry of Hamilton's equation.  In analogy with Eqs. (\ref{C_E})-(\ref{C_E+-}), we have that
\bea
&&\Theta \,  C_E =  C_E \qquad\qquad\qquad\ \; \mbox{for}\qquad U_{\rm min}<E<U_{\rm max} \; ,\\
&&\Theta \,  C_E^{(+)} =  C_E^{(-)}\neq  C_E^{(+)} \qquad \mbox{for}\qquad E>U_{\rm max} \; .
\eea
The simplicity of this example shows that the breaking of time-reversal symmetry is a common phenomenon arising by the selection of initial conditions. The fact is that the symmetry of the equations of motion does not imply the symmetry of all their solutions.

As a consequence, the invariant probability density (\ref{rho_time}) also breaks the symmetry if it is supported by one of the periodic orbits $ C_E^{(\pm)}$:
\be
\rho_{ C_E^{(+)}}(\Theta\Gamma) = \rho_{ C_E^{(-)}}(\Gamma) \neq \rho_{ C_E^{(+)}}(\Gamma)
\ee
for $E>U_{\rm max}$.  More generally, it is possible to define a continuous family of invariant densities supported by both periodic orbits according to
\be
\rho_{\lambda} = \lambda \, \rho_{ C_E^{(+)}} + (1-\lambda) \, \rho_{ C_E^{(-)}}
\ee
with $0\leq \lambda \leq 1$.  They satisfy the symmetry relation
\be
\rho_{\lambda}(\Theta\Gamma) = \rho_{1-\lambda}(\Gamma) 
\ee
and they break the time-reversal symmetry unless $\lambda = 1/2$.  In this latter case, we recover the microcanonical invariant density (\ref{rho_micro}), which is always symmetric under time reversal since it is defined in terms of the symmetric Hamiltonian function: $\rho_{\lambda=1/2}=\rho_E=\rho_E\circ\Theta$.

\vskip 0.3 cm

\begin{figure}[h]
\begin{center}
\includegraphics[scale=0.3]{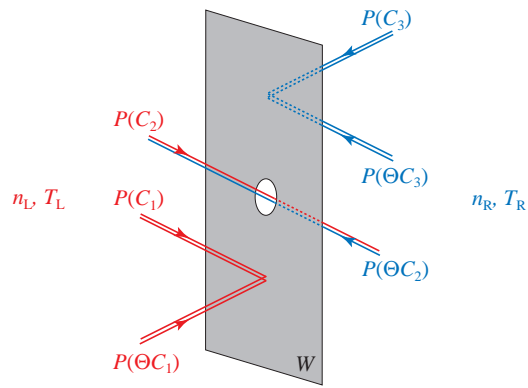}
\caption{Schematical representation of the effusion process in the physical position space ${\bf r}=(x,y,z)\in{\mathbb R}^3$. The wall $W$ separates the position space in two domains where the gas of noninteracting particles has different temperatures and densities.  The figure depicts three types of phase-space orbits $ C_i$ and their time reversal $\Theta C_i$ (with $i=1,2,3$).  They are here projected onto the position space by the projection $P:({\bf r},{\bf p})\in{\mathbb R}^6\to{\bf r}\in{\mathbb R}^3$.}
\label{fig2}
\end{center}
\end{figure}

\noindent{\bf 3. Effusion.}  In order to show that the previous considerations extend to situations relevant to nonequilibrium statistical mechanics, let us envisage the effusion process of an ideal gas of noninteracting particles through a small hole in a thin wall separating two infinite domains in the three-dimensional physical space \cite{GA11}.  The process is depicted in Fig.~\ref{fig2}.
In the domain on the left-hand side, the particles arrive from infinity with velocities and positions distributed according to a Maxwell-Boltzmann distribution at the temperature $T_{\rm L}$ and particle density $n_{\rm L}$.  Most of these particles undergo specular collisions on the wall and return to infinity in the same domain.  However, some particles cross the wall through the hole towards the other domain.  {\it Vice versa}, the domain on the right-hand side contains particles coming from the opposite infinity with velocities and positions distributed at the temperature $T_{\rm R}$ and particle density~$n_{\rm R}$.  Most of them remain in the same domain after one specular collision on the wall, while some of them cross the wall through the hole.  

The motion of every particle is ruled by the Hamiltonian flow $\Phi^t$ in the single-particle phase space ${\cal M}=\{\Gamma=({\bf r},{\bf p})\in({\mathbb R}^3-W)\otimes{\mathbb R}^3\}$ where $W$ denotes the thin wall with its hole.  The Hamiltonian function is given by $H={\bf p}^2/(2m)$, but the orbits undergo specular collisions on the wall.  Accordingly, the Hamiltonian flow is symplectic and time-reversal symmetric: $\Theta\circ\Phi^t\circ\Theta=\Phi^{-t}$.

The single-particle distribution function has the Maxwell-Boltzmann form:
\be
f(\Gamma)=f({\bf r},{\bf p}) = \frac{n_C}{(2\pi mk_{\rm B} T_C)^{3/2}} \, \exp\left(-\frac{{\bf p}^2}{2mk_{\rm B} T_C}\right)
\label{distrib_fn}
\ee
where the particle density $n_C$ and the temperature $T_C$ take the values associated with the orbit $C$ to which the phase-space point $\Gamma=({\bf r},{\bf p})$ belongs and corresponding to the domain from which the orbit is coming.  ($k_{\rm B}$ is Boltzmann's constant.)  For the three types of possible orbits depicted in Fig.~\ref{fig2}, these values are the following:
\bea
&& n_C=n_{\rm L}\; , \quad T_C=T_{\rm L} \qquad\,\mbox{for}\qquad \Gamma \in C_1\; , \ \Theta C_1 \; , \ C_2 \; ;\\
&& n_C=n_{\rm R}\; , \quad T_C=T_{\rm R} \qquad\mbox{for}\qquad \Gamma \in C_3\; , \ \Theta C_3 \; , \ \Theta C_2 \; .
\eea
The distribution function is invariant under the Hamiltonian flow, $f(\Phi^t\Gamma)=f(\Gamma)$.
Nevertheless, it is not symmetric under time reversal, $f(\Theta\Gamma)\neq f(\Gamma)$, for nonequilibrium constraints such that $n_{\rm L}\neq n_{\rm R}$ or $T_{\rm L}\neq T_{\rm R}$.

For the infinite-particle system, an invariant probability measure $\mu$ can be constructed as a Poisson suspension on the basis of the invariant distribution function (\ref{distrib_fn}), as explained in Ref.~\cite{CFS82}.  The so-constructed infinite dynamical system $(\Phi_{\infty}^t,{\mathcal M}^{\infty},\mu)$ is not symmetric under time reversal because $\mu\circ\Theta\neq \mu$ under nonequilibrium constraints,  although the flow $\Phi_{\infty}^t$ itself always is, as pointed out in Ref.~\cite{G98}.

\section{Symmetry relations for equilibrium steady states}
\label{Space}

In this section, we consider spin systems at equilibrium as described by the Hamiltonian function:
\be
H_B(\pmb{\sigma}) =  H_0(\pmb{\sigma}) - B M_N(\pmb{\sigma})
\ee
where $\pmb{\sigma}=(\sigma_1,\sigma_2,...,\sigma_N)$ is a configuration of the spins $\sigma_j=\pm 1$, $H_0(\pmb{\sigma})$ is a Hamiltonian that is symmetric under the group ${\mathbb Z}_2=\{1,\Pi\}$ with $\Pi\pmb{\sigma}=-\pmb{\sigma}$ and another group $G$ of spin permutations including translations of the type $T(\sigma_1,\sigma_2,\sigma_3...,\sigma_N)=(\sigma_N,\sigma_1,\sigma_2,...,\sigma_{N-1})$.  Moreover, $B$ denotes an external and uniform magnetic field and
\be
M_N(\pmb{\sigma}) = \sum_{j=1}^{N} \sigma_j
\label{M}
\ee
is the total magnetization.  Accordingly, the total Hamiltonian $H_B(\pmb{\sigma})$ is symmetric under the group $G$, but the ${\mathbb Z}_2$ symmetry is broken if $B\neq 0$.  We suppose that the system is in the Gibbsian equilibrium state
\be
\mu(\pmb{\sigma}) = \frac{1}{Z} \, {\rm e}^{-\beta H_B(\pmb{\sigma})}
\ee
with the inverse temperature $\beta=(k_{\rm B}T)^{-1}$ and the partition function $Z={\rm tr}\, {\rm e}^{-\beta H_B(\pmb{\sigma})}$.  This equilibrium state has the same symmetries as the Hamiltonian.
The fluctuations of the magnetization (\ref{M}) can be characterized in terms of large-deviation functions \cite{E85,T09}.

The main result is that, in the thermodynamic limit $N\to\infty$, the equilibrium probability distribution $P_B(M)$ of the magnetization (\ref{M}) obeys the symmetry relation:
\be
\frac{P_B(M)}{P_B(-M)} \simeq_{N\to\infty} {\rm e}^{2\beta BM} \; .
\label{sym_equil_P}
\ee
Introducing the generating function of the magnetization cumulants as
\be
Q_B(\lambda) \equiv \lim_{N\to\infty} -\frac{1}{N} \ln \langle {\rm e}^{-\lambda M_N}\rangle_B
\label{Q_B}
\ee
the symmetry relation (\ref{sym_equil_P}) yields the equivalent form:
\be
Q_B(\lambda) = Q_B(2\beta B-\lambda)
\label{sym_equil_Q}
\ee
because the large-deviation functions $P_B(M)$ and $Q_B(\lambda)$ are the Legendre transforms of each other.  Equations (\ref{sym_equil_P}) and (\ref{sym_equil_Q}) express the breaking of the ${\mathbb Z}_2$ symmetry by the external magnetic field. These results are established here below for two different models by using their large-deviation properties reported in Ref.~\cite{E85}.

\vskip 0.3 cm

\noindent{\bf 1. Ising model.}
The spins form a one-dimensional chain and their interaction energy is given by
\be
H_B(\pmb{\sigma}) = - J \sum_{j=1}^N \sigma_j \; \sigma_{j+1} - B \sum_{j=1}^N \sigma_j
\ee
with $\sigma_{N+1}=\sigma_1$.
This Hamiltonian function is symmetric under the group of translations $G=\{1,T,T^2,...,T^{N-1}\}$.
Now, the generating function (\ref{Q_B}) can be calculated in terms of the largest eigenvalue of the transfer matrix
\be
\hat V_{\lambda} = 
\left(\begin{array}{cc}
{\rm e}^{\beta J + \beta B-\lambda} & {\rm e}^{-\beta J}\\
{\rm e}^{-\beta J} & {\rm e}^{\beta J - \beta B+\lambda}
\end{array}
\right)
\ee
which obeys the symmetry relation
\be
\hat\pi \; \hat V_{\lambda} \; \hat\pi = \hat V_{2\beta B-\lambda} \qquad \mbox{with} \qquad 
\hat\pi =
\left(\begin{array}{cc}
0 & 1\\
1 & 0
\end{array}
\right) \; .
\ee
Accordingly, its eigenvalues have the symmetry $\lambda \to 2\beta B- \lambda$ and the relation (\ref{sym_equil_Q}) is satisfied. 

\vskip 0.3 cm

\noindent{\bf 2. Curie-Weiss model.}
Here, the Hamiltonian function can be written as
\be
H_B(\pmb{\sigma}) = - \frac{J}{2N} M_N(\pmb{\sigma})^2  - B \, M_N(\pmb{\sigma})
\ee
in terms of the magnetization (\ref{M}).  This Hamiltonian is invariant under the symmetric group $G={\rm Sym}\, N$, which contains subgroups similar to the translation group.  Moreover, the ${\mathbb Z}_2$ symmetry is broken if $B\neq 0$.  The number of spin configurations such that the magnetization takes the value $M_N=N-2n$ is equal to $g_n=N!/[n!(N-n)!]$ with $n=0,1,2,...,N$.  In the thermodynamic limit $N\to\infty$, the generating function (\ref{Q_B}) is given by
\be
Q_B(\lambda) = \lim_{N\to\infty} \frac{1}{N} \ln \frac{W_N(\beta B)}{W_N(\beta B-\lambda)}
\ee
with
\be
W_N(\xi) = \int_{-1}^{+1} dm \; \exp\left[N\left( \frac{\beta J}{2} \, m^2 + \xi \, m - \frac{1-m}{2}\, \ln\frac{1-m}{2}- \frac{1+m}{2}\,\ln\frac{1+m}{2}\right)\right]
\label{W_N}
\ee
where $m=1-2n/N$ stands for the magnetization per spin.  Now, the symmetry $W_N(\xi)=W_N(-\xi)$ of the partition function (\ref{W_N}) implies the symmetry relation (\ref{sym_equil_Q}), hence Eq.~(\ref{sym_equil_P}).  Spontaneous symmetry breaking only happens if $B=0$ and below the critical temperature of the paramagnetic-ferromagnetic transition where the magnetization probability distribution $P_B(M)$ is bimodal.

\section{Symmetry relations for nonequilibrium steady states}
\label{Time}

The symmetry relation (\ref{sym_equil_P}) is analogue to relationships established for nonequilibrium steady states such as the current fluctuation relation:
\be
\frac{P_{\bf A}({\bf J})}{P_{\bf A}(-{\bf J})} \simeq_{t\to\infty} \exp \left( {\bf A}\cdot{\bf J} \, t\right) 
\label{FTC}
\ee
which has been proved for all the fluctuating currents $\bf J$ flowing across some mesoscopic device under nonequilibrium constraints fixed by the thermodynamic forces, also called the affinities $\bf A$ \cite{AG07JSP}.  In terms of the cumulant generating function
\be
Q_{\bf A}(\pmb{\lambda}) = \lim_{t\to\infty} -\frac{1}{t}\ln \langle {\rm e}^{-\pmb{\lambda}\cdot{\bf J}_t}\rangle_{\bf A}
\ee
the relation (\ref{FTC}) implies the equivalent expression:
\be
Q_{\bf A}(\pmb{\lambda}) = Q_{\bf A}({\bf A}-\pmb{\lambda}) 
\label{FTC_Q}
\ee
which is similar to Eq.~(\ref{sym_equil_Q}).
Here, the relations (\ref{FTC}) and (\ref{FTC_Q}) express the breaking of the time-reversal symmetry in the nonequilibrium steady state driven by non-vanishing affinities $\bf A$.  The principle of detailed balancing is recovered at equilibrium for ${\bf A}=0$.  These relations have fundamental consequences on the linear and nonlinear response properties \cite{AG07JSM}. An example is given by the effusion process with the invariant probability measure constructed here above, as shown in Ref.~\cite{GA11}.

\section{Conclusions and perspectives}
\label{Conclusions}

In this paper, the analogy has been developed between different kinds of broken ${\mathbb Z}_2$ symmetries. In Section \ref{Hamilton}, simple Hamiltonian flows already illustrate these symmetry-breaking phenomena induced by the selection of initial conditions and their profound similarities.  In Sections \ref{Space} and \ref{Time}, this analogy has been further emphasized. On the one hand, Eqs.~(\ref{sym_equil_P}) and (\ref{sym_equil_Q}) have been established for two equilibrium models where the space-inversion symmetry is broken by an external magnetic field.  On the other hand, the similar relationships (\ref{FTC}) and (\ref{FTC_Q}) characterize the current fluctuations in nonequilibrium steady states where the time-reversal symmetry is broken by some external drivings $\bf A$.  These results express the effects of the broken ${\mathbb Z}_2$ symmetry on the fluctuations of the relevant quantities, showing the generality of such considerations, which concern other ${\mathbb Z}_2$ symmetries as well \cite{J10}.

\begin{acknowledgments}
Financial support from the European Science Foundation and the Belgian Federal Government (IAP project ``NOSY") is kindly acknowledged.
\end{acknowledgments}


\end{document}